\documentclass[aps,prd,preprint,showpacs,showkeys,superscriptaddress,nofootinbib]{revtex4-1}

\usepackage{booktabs}
\usepackage{graphicx}
\usepackage{xcolor}
\usepackage{amsmath}
\usepackage{hyperref}
\usepackage{bm}
\usepackage{mathtools}
\usepackage{amssymb}
\usepackage{lipsum}
\usepackage{epstopdf}
\usepackage{slashed}
\usepackage{multirow}
\usepackage{appendix}

\AtBeginDocument{
\heavyrulewidth=.08em
\lightrulewidth=.05em
\cmidrulewidth=.03em
\belowrulesep=.65ex
\belowbottomsep=0pt
\aboverulesep=.4ex
\abovetopsep=0pt
\cmidrulesep=\doublerulesep
\cmidrulekern=.5em
\defaultaddspace=.5em
}

\begin{document}
\def\qq{\langle \bar q q \rangle}
\def\uu{\langle \bar u u \rangle}
\def\dd{\langle \bar d d \rangle}
\def\sp{\langle \bar s s \rangle}
\def\GG{\langle g_s^2 GG \rangle}
\def\Tr{\mbox{Tr}}

\def\ds{\displaystyle}
\def\beq{\begin{equation}}
\def\eeq{\end{equation}}
\def\bea{\begin{eqnarray}}
\def\eea{\end{eqnarray}}
\def\beeq{\begin{eqnarray}}
\def\eeeq{\end{eqnarray}}
\def\ve{\vert}
\def\vel{\left|}
\def\ver{\right|}
\def\nnb{\nonumber}
\def\ga{\left(}
\def\dr{\right)}
\def\aga{\left\{}
\def\adr{\right\}}
\def\lla{\left<}
\def\rra{\right>}
\def\rar{\rightarrow}
\def\nnb{\nonumber}
\def\la{\langle}
\def\ra{\rangle}
\def\ba{\begin{array}}
\def\ea{\end{array}}
\def\tr{\mbox{Tr}}
\def\ssp{{\Sigma^{*+}}}
\def\sso{{\Sigma^{*0}}}
\def\ssm{{\Sigma^{*-}}}
\def\xis0{{\Xi^{*0}}}
\def\xism{{\Xi^{*-}}}

\def\qs{\la \bar s s \ra}
\def\qu{\la \bar u u \ra}
\def\qd{\la \bar d d \ra}

\def\gGgG{\la g^2 G^2 \ra}
\def\q{\gamma_5 \not\!q}
\def\x{\gamma_5 \not\!x}
\def\g5{\gamma_5}
\def\sb{S_Q^{cf}}
\def\sd{S_d^{be}}
\def\su{S_u^{ad}}
\def\sbp{{S}_Q^{'cf}}
\def\sdp{{S}_d^{'be}}
\def\sup{{S}_u^{'ad}}
\def\ssp{{S}_s^{'??}}

\def\sig{\sigma_{\mu \nu} \gamma_5 p^\mu q^\nu}
\def\fo{f_0(\frac{s_0}{M^2})}
\def\ffi{f_1(\frac{s_0}{M^2})}
\def\fii{f_2(\frac{s_0}{M^2})}
\def\O{{\cal O}}
\def\sl{{\Sigma^0 \Lambda}}
\def\es{\!\!\! &~=~& \!\!\!}
\def\ap{\!\!\! &\approx& \!\!\!}
\def\ar{&+& \!\!\!}
\def\ek{&-& \!\!\!}
\def\kek{\!\!\!&-& \!\!\!}
\def\cp{&\times& \!\!\!}
\def\se{\!\!\! &\simeq& \!\!\!}
\def\eqv{&\equiv& \!\!\!}
\def\kpm{&\pm& \!\!\!}
\def\kmp{&\mp& \!\!\!}
\def\mcdot{\!\cdot\!}
\def\erar{&\rightarrow&}


\title{Multipole Moments of Heavy Vector and Axial-Vector Mesons in QCD}

\author{T.~M.~Aliev}
\email{taliev@metu.edu.tr}
\affiliation{Department of Physics, Middle East Technical University, Ankara, 06800, Turkey}

\author{S.~Bilmis}
\email{sbilmis@metu.edu.tr}
\affiliation{Department of Physics, Middle East Technical University, Ankara, 06800, Turkey}

\author{M.~Savci}
\email{savci@metu.edu.tr}
\affiliation{Department of Physics, Middle East Technical University, Ankara, 06800, Turkey}

\date{\today}

\begin{abstract}
The magnetic and quadrupole moments of the vector and axial-vector mesons containing heavy quark are estimated within the light cone sum rules method. Our predictions on magnetic moments for the vector mesons are compared with the results obtained by other approaches.
\end{abstract}

\maketitle

\section{Introduction}
A study of electromagnetic properties of hadrons plays a crucial role in understanding their inner structure. The magnetic moments are one of the fundamental characteristics of hadrons. The magnetic moments of hadrons are related to their magnetic form factors; more precisely, the magnetic moment is equal to the magnetic form factor at zero momentum square.
The magnetic moments of the mesons have not received much interest compared to the baryons except $\rho$-meson, which have been intensively studied in the literature within different approaches \cite{2009PhLB..678..470A,2008PhRvD..78i4502L,2007PhRvD..75i4504H,2003JHEP...12..061S,2013PhRvD..87g4012B,2008PhRvC..77b5203B,2002PhRvD..65k6001B,2004PhRvD..70e3015C,1999PhRvC..59.1743H,2015ChPhC..39k3103L,2014PhLB..730..115D}.
The magnetic moments of $K^*$ mesons have also been investigated in several works \cite{2009PhLB..678..470A,2007PhRvD..75i4504H,2008PhRvD..78i4502L,2013PhRvD..87g4012B,2008PhRvC..77b5203B,1999PhRvC..59.1743H,2015ChPhC..39k3103L}. On the other hand, the magnetic moments of heavy mesons have been calculated only in few works~\cite{2015ChPhC..39k3103L,1980PhRvD..22..773B,2003NuPhA.714..183L}. In the face of this situation, it is timely to study the magnetic moments of heavy vector and axial-vector mesons. It is challenging to measure the magnetic moments of the vector mesons directly since their lifetimes are very short. Even though the indirect measurement is possible~\cite{2014IJMPS..3560463T}, however, it has large uncertainty. It should be noted that with the help of the magnetic dipole transitions $M1$, in which there exists many experimental data, it is possible to determine the transition magnetic moments of heavy mesons. There are lots of theoretical works, such as quark model~\cite{1993PhLB..316..555C,1996PhRvD..53.1349J}, non-relativistic QCD~\cite{2006PhRvD..73e4005B} the quark potential model~\cite{1995PhRvD..51.3613G,1999PhRvD..60g4006F,2005PhRvD..72e4026B,2007PhLB..650..159L}, various relativistic models, the bag model~\cite{1985PhRvD..31.1081W,1993NuPhA.559..579Z,Hiorth1999}, the light front model~\cite{1994PhLB..336..113O,1995PhLB..359....1C,2007PhRvD..75g3016C}, the Bethe-Salpeter equation~\cite{2000EPJA....9...73K,2000NuPhA.674..141L}, QCD sum rules~\cite{1980PhLB...90..460K,1996PhLB..368..163D,1994PhLB..334..169A,1996PhRvD..54..857A,1980ZPhyC...4..345S,1984ZPhyC..26..275A,1985NuPhB.260...61B,1994PhLB..329..123A,Colangelo:2005hv}, lattice QCD~\cite{2011EPJC...71.1734B,2012PhRvD..86i4501D}, chiral model~\cite{1992PhLB..296..408C,1992PhLB..296..415A,2003PhRvD..68e4024B,2005EPJC...39...27H,2015PhRvD..91a4010C}, Nambu-Jona-Lasino model, the dispersion approach, etc. devoted to this subject.

In present work, we calculate the magnetic moments of heavy vector and axial-vector mesons within light cone QCD sum rules (LCSR)~\cite{Braun:1997kw}. The calculation of the multipole moments for the axial-vector mesons is performed for the first time.

The paper is organized as follows. In section II, we construct the LCSR for multipole moments of heavy vector and axial-vector mesons. The following section is devoted to the numerical analysis of the sum rules for the multipole moments of heavy vector and axial-vector mesons. In this section, the obtained results are also compared with predictions of other approaches in the literature.  
The last section contains a summary and discussions.
\section{Light cone sum rules for multipole moments}
The LCSR for multipole moments of vector (axial-vector) heavy mesons can be obtained by considering the following correlation function
\begin{equation}
  \label{eq:1}
  \Pi^{\mu \alpha \nu} = i^2 \int d^4 x \int d^4 y e^{i p x + iqy} \langle 0 | T \{ J^{\mu} (x) j^\alpha_{el}(y) J^{\nu^\dag} (0) \} |0 \rangle, 
\end{equation}
where $J^\mu(x) = \bar{q}^a(x) \gamma_\mu Q^a(x)$ is the interpolating current with the quantum numbers of a heavy vector meson and $a$ is the color index. The interpolating current for axial-vector mesons can be obtained from $J^{\mu}(x)$ with simple replacement $\gamma^\mu \rightarrow \gamma^\mu \gamma^5$. The current $j^\alpha_{el}(y) = e_q \bar{q} \gamma^\alpha q + e_Q \bar{Q} \gamma^\alpha Q$ is the electromagnetic current, $e_q$ and $e_Q$ are the electric charges of the light and heavy mesons, respectively.

The general strategy of QCD sum rules is that the correlation function has to be calculated in different kinematical domains. In one domain, it is saturated by the corresponding heavy vector (axial-vector) mesons, i.e. $p^2 \simeq  m_{V_Q}^2~(m_{A_Q}^2) (\text{hadronic part})$. In the other domain, where $p^2 \ll 0$, $(p + q)^2 \ll 0$, the calculation is performed by using the operator product expansion (OPE) in terms of the photon distribution amplitudes (DA's) with an increasing twist. Matching the results of these representations, one can get the desired sum rules for the relevant physical quantities.

The hadronic part of the correlation function can be obtained by inserting a complete set of states carrying the same quantum numbers of the interpolating currents. Isolating the ground state vector mesons, we have
\begin{equation}
  \label{eq:6}
  \Pi_{\mu \nu \alpha}(p,q) = \frac{ \langle 0 | J_\mu | i(p) \rangle \langle i(p) | j^{el}_\alpha(y) | f(p^\prime) \rangle \langle f(p^\prime) | J_\nu^+ | 0 \rangle + ...}{(p^2 - m_i^2)({p^\prime}^2 - m_i^2)}~.
\end{equation}

The matrix element $\langle 0 | J_\mu | i (p) \rangle $ in Eq.~\eqref{eq:6} is defined as
\begin{equation}
  \label{eq:7}
  \langle 0 | J_\mu | i(p) \rangle = f_i m_i \epsilon_\mu~,
\end{equation}
where $f_i$ is the leptonic decay constant of the corresponding heavy vector mesons and $\epsilon_\mu$ is its polarization vector. Using the parity and time-reversal invariance of the electromagnetic interaction the matrix element of the electromagnetic current between two vector (axial-vector) mesons is described in terms of three form-factors as~\cite{PhysRevD.46.2141} :

\begin{equation}
  \label{eq:8}
  \begin{split}
    \langle f (p^\prime, \epsilon^\prime) | j_{\alpha}^{el} | i (p, \epsilon^r) \rangle =& - (\epsilon^r)^\rho (\epsilon ^{r^\prime})^\beta \times \\
    & \big\{ G_1(Q^2) g_{\rho \beta} (p^\prime +p)_\alpha + G_2(Q^2) (q_\rho g_{\alpha \beta} - q_\beta g_{\alpha \rho})+ \frac{G_3(Q^2)}{2m_i^2} q_\rho q_\beta (p+p^\prime)_\alpha \big\}~,
  \end{split}
\end{equation}
where  $G_i$ are the form factors and $Q^2 = -q^2$. The form factors $G_i (Q^2)$ are related to charge, magnetic and quadrupole multipole form factors in the following way~\cite{PhysRevD.46.2141}
\begin{equation}
  \label{eq:3}
  \begin{split}
    F_c(Q^2) &= G(Q^2) + \frac{2}{3}\eta F_D(Q^2), \\
    F_M(Q^2) &= G_2(Q^2), \\
    F_D(Q^2) &= G_1(Q^2) - G_2(Q^2) + (1+\eta)G_3(Q^2), 
  \end{split}
\end{equation}
where $\eta = Q^2/ 4m_i^2$, $F_C(Q^2),~F_M(Q^2)$ and $F_D(Q^2)$ are the charge, magnetic and quadrupole form factors, respectively. The value of $F_c(Q^2)$, $F_M(Q^2)$, and $F_D(Q^2)$ at $Q^2 =-q^2 = 0$ point gives the charge, magnetic moment $\mu$ and quadrupole moment $D$ of the vector (axial-vector) mesons.

Substituting Eqs. \eqref{eq:7}, \eqref{eq:8} and \eqref{eq:3} into Eq. \eqref{eq:6}, and performing summation over the spins of vector mesons  for the hadronic part of the correlation function, we get
\begin{equation}
  \label{eq:2}
  \begin{split}
    \Pi_{\mu \alpha \nu} \epsilon^{\alpha}_\gamma  = & ~ f_i^2 m_i^2 \frac{\epsilon^\alpha_\gamma}{ (m_i^2 - p^2) (m_i^2 - (p+q)^2 )} \times \\
     \big\{ & 2 F_c(0) p_\alpha (g_{\mu \nu} - \frac{p_\mu p_\nu}{m_i^2} - \frac{p_\mu q_\nu}{m_i^2}) \\
    &+ F_{M}(0) \big[ q_\mu g_{\nu \alpha} - q_\nu g_{\mu \alpha} - \frac{p_\alpha}{m_i^2} (p_\mu q_\nu - p_\nu q_\mu)\big] \\
    &- \big[ F_c(0) + F_D(0) \big] \frac{p_\alpha}{m_i^2} q_\nu q_\mu \big\},
  \end{split}
\end{equation}
where $\epsilon_\gamma$ is the photon polarization vector. To derive this expression, the transversality condition $q \epsilon = 0$ is used.

Now let us turn our attention to the calculation of Eq.\eqref{eq:1} from the OPE side. By introducing the electromagnetic background field of a plane wave
\begin{equation}
  \label{eq:4}
  F_{\mu\nu} = i (e_\nu^\gamma q_\mu - e_\mu^\gamma q_\nu) e^{i q x}~,
\end{equation}
the correlation function can be written in the following way;
\begin{equation}
  \label{eq:5}
  \Pi_{\mu \alpha \nu} \epsilon^\alpha_\gamma = i \int d^4 x e^{i p x} \langle 0 | T \{J_\mu (x) J_\nu^{+}(0) \} |0 \rangle _F~.
\end{equation}

In this expression, subscript $F$ means that the vacuum expectation value is evaluated in the presence of the background field $F_{\mu \nu}$. The correlation function given in Eq.~\eqref{eq:1} can be obtained from Eq.~\eqref{eq:5} by expanding it in linear powers of $F_{\mu \nu}$. More details about the background field method is given in two excellent reviews~\cite{Rohrwild:2007yt,Ball:2002ps}.

Using the explicit expressions of the interpolating currents and applying the Wick theorem for the correlation function, we obtain
\begin{equation}
  \label{eq:13}
  \Pi_{\mu \alpha \nu} \epsilon^\alpha_\gamma = i \int d^4x e^{ipx} \langle 0 | S_Q(x) \gamma_\mu S(-x) \gamma_\nu |0 \rangle_F~.
\end{equation}
From this expression, it follows that to calculate the correlation function in the deep Euclidean domain, it is necessary to know the explicit expressions of the light and heavy quark propagators in the presence of the background gluonic and electromagnetic fields. The expressions of these propagators are obtained in~\cite{BALITSKY1989541,PhysRevD.51.6177}.
\begin{equation}
  \label{eq:11}
  \begin{split}
    S(x) &= \frac{i \slashed{x}}{2 \pi^2 x^4} - \frac{m_q}{4 \pi^2 x^2}
    - \frac{i g_s}{16\pi^2 x^2} \int_0^1 du \big\{ \bar{u} \slashed{x} \sigma_{\alpha \beta} + u \sigma_{\alpha \beta} \slashed{x} \big\} G^{\alpha \beta}(ux) \\
        & \frac{-i e_q}{16 \pi^2 x^2} \int_0^1 du \big\{ \bar{u} \slashed{x} \sigma_{\alpha \beta} + u \sigma_{\alpha \beta} \slashed{x} \big\} F^{\alpha \beta}(ux) \\
    &- i g_s \int_0^1 du \{ - \frac{i m_q}{32 \pi^2} G_{\mu \nu }(ux) \sigma^{\mu \nu} \ln{(- \frac{x^2 \Lambda^2}{u}+ 2 \gamma_E)} \} \\
    &- i e_q \int_0^1 du \{ - \frac{i m_q}{32 \pi^2} F_{\mu \nu }(ux) \sigma^{\mu \nu} \ln{(- \frac{x^2 \Lambda^2}{u}+ 2 \gamma_E)} \}    
  \end{split}
\end{equation}
\begin{equation}
  \label{eq:12}
  \begin{split}
    S_Q(x) &= \int \frac{d^4k}{2 \pi^4 i} e^{-ikx} \frac{\slashed{k}+m_Q}{m_Q^2 - k^2} \\
    &- i g_s \int \frac{d^4k}{2 \pi^4 i} e^{-ikx} \int_{0}^{1} du \big\{ \frac{\slashed{k}+m_Q}{2(m_Q^2 - k^2)^2} G^{\alpha \beta} \sigma_{\alpha \beta} + \frac{u x_\alpha}{m_Q^2 - k^2} G^{\alpha \beta}(ux) \gamma_\beta \big\} \\
    &- ie_Q \int \frac{d^4k}{2 \pi^4 i} e^{-ikx} \int_{0}^{1} du \big\{ \frac{\slashed{k}+m_Q}{2(m_Q^2 - k^2)} F^{\alpha \beta}(u x) \sigma_{\alpha \beta} +\frac{u x_\alpha}{m_Q^2 - k^2}F^{\alpha \beta} \gamma_\beta \big\}
  \end{split}
\end{equation}
where $\bar{u} = 1-u$, $\Lambda = (0.5 \pm 0.1)~\rm{GeV}$~\cite{Chetyrkin:2007vm}, is the scale parameter separating the perturbative and nonperturbative domains, and $\gamma_E = 0.577$ is the Euler constant. Note that the four quark particle $\bar{q}q \bar{q} q$ and $\bar{q} G^2 q$ operators contributions are small and is not presented in Eq.~\ref{eq:11}~\cite{BALITSKY1989541,Braun:1989iv}.

In light-cone sum rules, the nonperturbative contribution appears when a photon is emitted at long distances. To obtain these contributions, it is necessary to expand the quark propagator near the light cone $x^2 = 0$. In this case, the following matrix elements of non-local operators in the presence of the external background field needs to be evaluated
\begin{equation}
  \label{eq:9}
  \begin{split}
    & \langle 0 | \bar{q}(x) \Gamma q(0) | 0 \rangle _F~, \\
    & \langle 0 | \bar{q}(x) \Gamma G_{\alpha \beta} q(0) |0 \rangle _F~, \\
    & \langle 0 | \bar{q}(x) \Gamma F_{\mu \nu} q(0) |0 \rangle _F~, \\
  \end{split}
\end{equation}
where $\Gamma$ is arbitrary Dirac matrices.
These matrix elements are described by photon distribution amplitudes, which were determined in~\cite{Ball:2002ps} and presented in Appendix A for completeness.

From Eq.~\eqref{eq:2}, it follows that we have numerous structures which can be used to calculate the magnetic and quadrupole moments of heavy vector (axial vector) mesons. We adopt the structures $(p \epsilon) p_\mu p_\nu$, $(p \epsilon) p_\nu q_\mu$ to determine $F_c(0)$, $F_M(0)$, and $F_c(0) + F_D(0)$. The Choice of these structures is dictated by the fact that they contain the maximal number of momenta, which exhibits good convergence in general. The theoretical part of the correlation function can be obtained from Eq.\eqref{eq:13} by substituting the explicit expressions of the heavy and light quark propagators and the photon DAs. Performing integration over $x$, the expression of the correlation function in the momentum representation can be obtained. Matching these two expressions of the correlation function via dispersion relation and performing doubly Borel transformations on the $-p^2$ and $-(p + q)^2$ in order to suppress the contributions of higher states and continuum, we get the desired sum rules for the multipole form factors. Note that the higher states contributions are taken into account by using the quark-hadron duality ansatz.

In result, we get the following sum rules for the charge $F_c(0)$, magnetic moment $F_M(0)$ and the sum of charge and quadrupole moment form factors at $Q^2 = 0$ point
\begin{equation}
  \label{eq:10}
  \begin{split}
    F_c(0) & = -\frac{1}{2 f_i^2} e^{m_i^2/M^2}~\Pi_1^{(\pm)}~, \\
    F_M(0) & = -\frac{1}{f_i^2} e^{m_i^2/M^2}~\Pi_2^{(\pm)}~, \\
    F_c(0) + F_D(0) & = -\frac{1}{f_i^2} e^{m_i^2/M^2}~\Pi_3^{(\pm)}~. \\
  \end{split}
\end{equation}
Explicit expressions of $\Pi_1^{(\pm)}$, $\Pi_2^{(\pm)}$, and $\Pi_3^{(\pm)}$ are presented in Appendix B. The upper (lower) sign corresponds to vector (axial-vector) mesons. Moreover, we denote $D_1$ and $D_{s1}$ axial-vector mesons with mass $2420~\rm{MeV}$ and $2460~\rm{MeV}$, respecctively.

\section{Numerical Analysis}
This section is devoted to the numerical analysis of the sum rules for the multipole moments of the heavy vector (axial-vector mesons). The values of the input parameters entering the sum rules are presented in Table~\ref{tab:1}.
\begin{table}[h]
 \renewcommand{\arraystretch}{1.05}
\setlength{\tabcolsep}{10pt}
    \centering
    \begin{tabular}{lllll}
        \toprule
$ \qq(1~GeV)$          & $(-0.246^{+0.028}_{-0.019})^3~GeV^3$~\cite{PhysRevD.80.114005}  \\
 $\sp(1~GeV)$          & $0.8 \times \langle \bar{q}q \rangle$~\cite{PhysRevD.80.114005}                        \\
$m_0^2$                & $(0.8   \pm 0.2)~GeV^2$~\cite{Belyaev:1982sa}                                     \\
$m_s(2~GeV)$           & $(96^{+8}_{-4} \times 10^{-3})~GeV$~\cite{PhysRevD.98.030001}   \\ 
$f_{{\cal D}^*}$       & $(0.263 \pm 0.021)~GeV $~\cite{Wang:2015mxa}                                   \\
$f_{{{\cal D}_{s}^*}}$ & $(0.308  \pm 0.021)~GeV$~~\cite{Wang:2015mxa}                                    \\
$f_{B^\ast}$           & $ \left(0.196^{+0.028}_{-0.027}\right)~GeV$~\cite{Wang:2015mxa} \\   
$f_{B_s^\ast}$         & $ (0.255 \pm 0.019)~GeV$~\cite{Wang:2015mxa}                    \\
$f_{{\cal D}_1}$       & $(0.332 \pm 0.018~)~GeV$~ ~\cite{Wang:2015mxa}                                    \\
$f_{{{\cal D}_{s1}}}$ & $(0.245  \pm 0.017)~GeV$~ ~\cite{Wang:2015mxa}                                    \\
$f_{B_1}$              & $ (0.335 \pm 0.018)~GeV$~\cite{Wang:2015mxa}                    \\
$f_{{B_{s1}}}$        & $ (0.348 \pm 0.018)~GeV$~\cite{Wang:2015mxa}                    \\
$\chi (1~GeV)$         & $ -(2.85 \pm 0.5)~GeV^{-2}$~\cite{Rohrwild:2007yt}                 \\
$f_{3 \gamma}$         & $ - 0.0039~GeV^{-2}$~\cite{Ball:2002ps}                        \\    
        \bottomrule
    \end{tabular}
    \caption{The values of the input parameters.}
    \label{tab:1}
  \end{table}
In this study, we use the $\overline{MS}$ mass, $m_c~(m_c) = (1.275 \pm 0.035~\rm{GeV})$, $m_b~(m_b) = (4.18 \pm 0.03~\rm{GeV})$  and take into account the scale dependence of the $\overline{MS}$ masses coming from renormalization group equation
  \begin{equation}
    \label{eq:14}
    \begin{split}
      m_b(\mu) = m_b (m_b) \big( \frac{\alpha_s (\mu)}{\alpha_s(m_b)} \big)^{12/23}~, \\
      m_c(\mu) = m_c(m_c)  \big( \frac{\alpha_s (\mu)}{\alpha_s(m_c)} \big)^{12/25}~.
    \end{split}
  \end{equation}
\begin{table}[h]
\renewcommand{\arraystretch}{1.1}
\setlength{\tabcolsep}{10pt}
\centering
  \begin{tabular}{ccc}
\toprule
    Mesons      & $M^2~(\rm{GeV^2})$  & $s_0~(\rm{GeV^2})$ \\
    \midrule
    $B^*$       & $(11 \pm 3)$        & $(35 \pm 1)$ \\
    $B_s^*$     & $(12 \pm 3)$        & $(37 \pm 1)$ \\
    $B_1$     & $(13 \pm 2)$        & $(42 \pm 1)$ \\
    $B_{s1}$  & $(14 \pm 3)$        & $(43 \pm 1)$ \\
    $D^*$      & $(4.5 \pm 1.5)$     & $(6.5 \pm 0.5)$ \\
    $D_s^*$    & $(4.5 \pm 1.5)$     & $(7.5 \pm 0.5)$ \\
    $D_1$      & $(5 \pm 2)$       & $(8.5 \pm 0.5)$ \\
    $D_{s1}$    & $(5 \pm 2)$      & $(9.5 \pm 0.5)$ \\
    \bottomrule
  \end{tabular}
  \caption{Working region of $M^2$ and $s_0$ parameters are shown.}
  \label{tab:3}
\end{table}
Besides the input parameters that are presented in Tables~\ref{tab:1}, sum rules contain two more extra parameters, namely, the continuum threshold $s_0$ and the Borel mass parameter $M^2$. The domain of $M^2$ is determined by demanding the standard criteria, namely, both power corrections and continuum contributions should be sufficiently suppressed. The continuum threshold is determined from the condition that the sum rules should reproduce the mass of the ground state mass with $10\%$ accuracy. These conditions are fulfilled in the regions of $M^2$, and $s_0$ presented in Table~\ref{tab:3}.
Having specified all input parameters, we are ready to calculate the numerical values of the magnetic and quadrupole moments, i.e. corresponding form-factors at $q^2 = 0$  point of all considered vector and axial-vector mesons.

\begin{figure}[hbt!]
  \centering
  \includegraphics[scale=0.60]{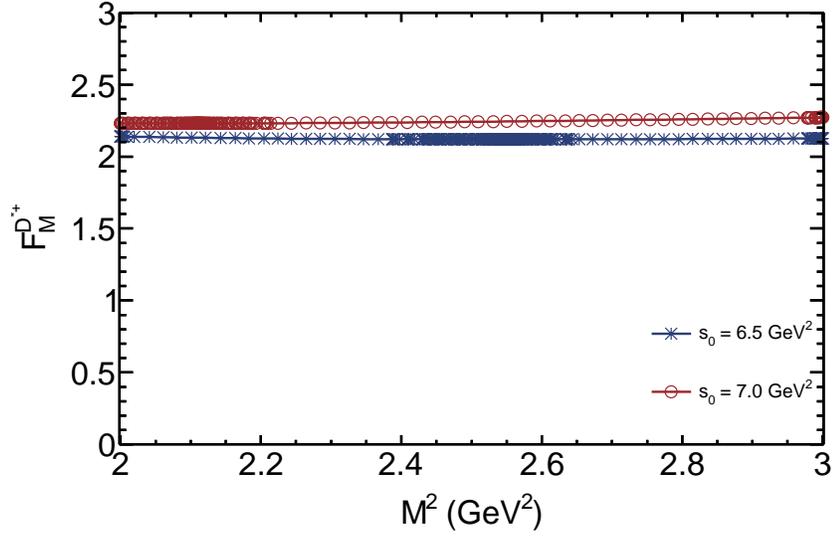}
  \caption{The dependency of the magnetic moment of $D^{*+}$ meson on $M^2$ at two fixed values of $s_0$.}
  \label{fig:1}
\end{figure}

\begin{figure}[hbt!]
  \centering
  \includegraphics[scale=0.60]{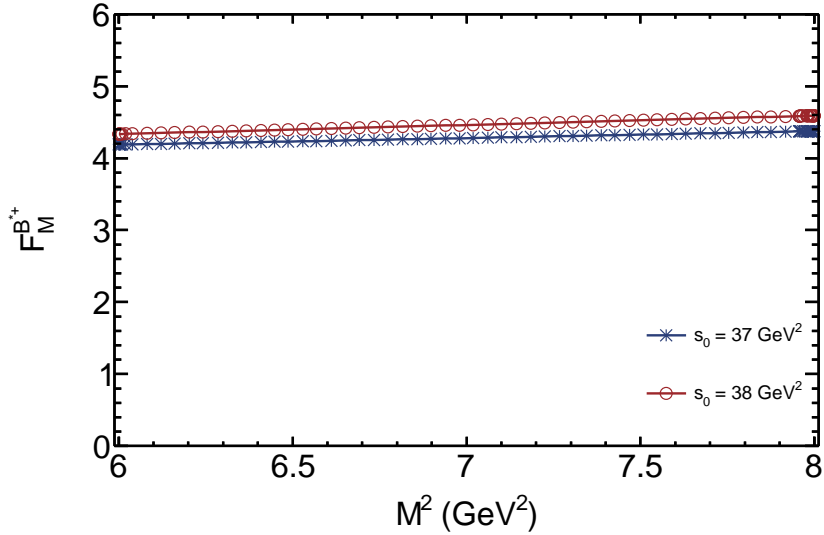}
  \caption{The same as in Fig.1, but for $B^{*+}$ meson.}
  \label{fig:2}
\end{figure}

\begin{figure}[hbt!]
  \centering
  \includegraphics[scale=0.60]{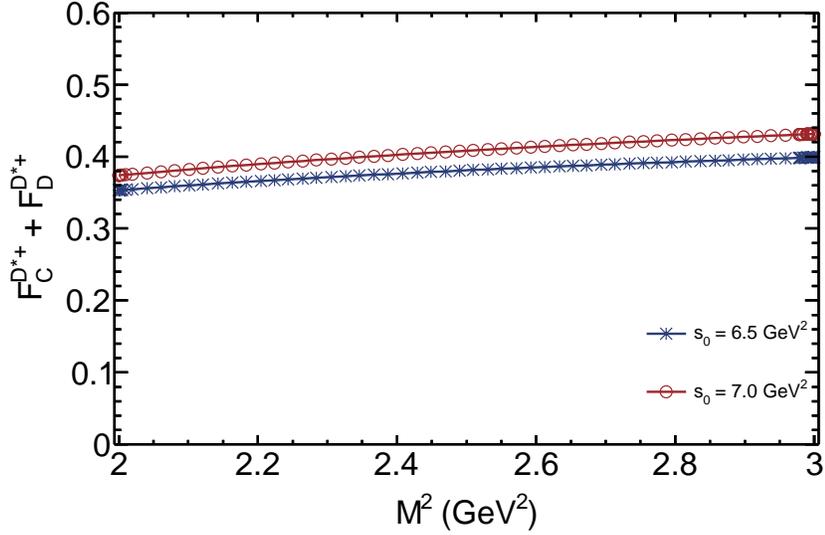}
  \caption{The dependency of $F_C^{D^{*+}} + F_D^{D^{*+}}$ on $M^2$ for $D^{*+}$ meson.}
  \label{fig:3}
\end{figure}

In Fig.~\ref{fig:1} and Fig.~\ref{fig:2}, we presented the dependency of $F_M^{D^{*+}}$ and $F_M^{B^{*+}}$ on $M^2$ at two fixed values of continuum threshold, respectively, for illustration. From these figures, we observe good stability of $F_M^{D^{*+}}$ and $F_M^{B^{*+}}$ to the variation of $M^2$. In Fig.~\ref{fig:3}, we depict the dependence of $F_C^{D^{*+}} + F_D^{D^{*+}}$ on $M^2$ at two fixed values of $s_0$. Similar to the magnetic momentum case $F_M(0)$, the  $F_C^{D^{*+}} +F_D^{D^{*+}}$ shows weak dependency to the variation of $M^2$. Performing similar calculations for  all vector and axial vector mesons considered, we get the values of $F_M(0)$ and $F_D(0)$ presented in Table~\ref{tab:4} and \ref{tab:5}, respectively.

The uncertainties result from the variation of Borel parameter $M^2$, continuum threshold $s_0$ as well as from uncertainties in input parameters. All uncertainties are taken quadratically. Moreover, for completeness, in Table~\ref{tab:4}, we also present the predictions on the magnetic moment obtained from non-relativistic quark model~\cite{PhysRevD.80.114005}, Nambu-Jona-Lasino model~\cite{2015ChPhC..39k3103L}, bag model~\cite{1980PhRvD..22..773B} expanded bag model~\cite{Simonis:2016pnh}, and chiral perturbation theory (ChPT)~\cite{PhysRevD.100.016019}.

\begin{table}
 \renewcommand{\arraystretch}{1.05}
  \setlength{\tabcolsep}{10pt}
  \centering
  \begin{tabular}{cccccccc}
    \toprule
Particle       & Our                & NJL                        & NR       & Bag                        & Extended-Bag           & ChPT                        \\ 
               &                    & \cite{2015ChPhC..39k3103L} & \cite{PhysRevD.98.030001}         & \cite{1980PhRvD..22..773B} & \cite{Simonis:2016pnh} & \cite{PhysRevD.100.016019} \\ 
\midrule
  $D^{*0}$     & $(0.30 \pm 0.04)$  & $\dotsi $                  & $-1.47$  & $-0.89$                    & $-1.28$                & $1.48^{+0.22}_{-0.38}$     \\
  $D^{*+}$     & $(1.16 \pm 0.08)$  & $1.16$                     & $1.32$   & $1.17$                     & $1.13$                 & $1.62^{+0.24}_{-0.08}$     \\
  $D_{s}^{*}$ & $(1.00 \pm 0.14) $ & $0.98$                     & $1.00$   & $1.03$                     & $0.93$                 & $0.69^{+0.22}_{-0.10}$     \\
  $B^{*+}$     & $(0.90 \pm 0.19)$  & $1.47$                     & $1.92$   & $1.54$                     & $1.56$                 & $1.77^{+0.25}_{-0.30}$     \\
  $B^{*0}$     & $-(0.21 \pm 0.04)$ & $\dotsi $                  & $-0.87$  & $-0.64$                    & $-0.69$                & $-0.92^{+0.15}_{-0.11}$    \\
  $B_{s}^{*}$ & $-(0.17 \pm 0.02)$ & $\dotsi $                  & $-0.55$  & $-0.47$                    & $-0.51$                & $-0.27^{+0.13}_{-0.10}$    \\
  $D_{1}^{0}$  & $(0.18 \pm 0.04$)  & $\dotsi$                   & $\dotsi$ & $\dotsi$                   & $\dotsi$               & $\dotsi$                   \\
  $D_{1}^{+}$  & $(0.90 \pm 0.08$)  & $\dotsi$                   & $\dotsi$ & $\dotsi$                   & $\dotsi$               & $\dotsi$                   \\
  $D_{s1}^{}$ & $(0.87 \pm 0.08)$  & $\dotsi$                   & $\dotsi$ & $\dotsi$                   & $\dotsi$               & $\dotsi$                   \\
  $B_{1}^{0}$  & $(0.14 \pm 0.08)$  & $\dotsi$                   & $\dotsi$ & $\dotsi$                   & $\dotsi$               & $\dotsi$                   \\
  $B_{1}^{+}$  & $(0.60 \pm 0.07)$  & $\dotsi$                   & $\dotsi$ & $\dotsi$                   & $\dotsi$               & $\dotsi$                   \\
  $B_{s1}^{}$ & $(0.13 \pm 0.09)$  & $\dotsi$                   & $\dotsi$ & $\dotsi$                   & $\dotsi$               & $\dotsi$                   \\
    \bottomrule
  \end{tabular}
  \caption{Magnetic moments (in nuclear magneton) of heavy vector and axial vector mesons.}
  \label{tab:4}
\end{table}

\begin{table}
 \renewcommand{\arraystretch}{1.05}
\setlength{\tabcolsep}{10pt}
    \centering
    \begin{tabular}{cc}
      \toprule
      \multicolumn{2}{c}{$F_D$ (in $e/m_i^2$ unit)} \\
        \midrule 
  $D^{*0}$           &  $(0.25 \pm 0.05)$ \\  
  $D^{*+}$           &  $-(0.64 \pm 0.02)$ \\  
  $D_{s}^{*}$        &  $-(0.60 \pm 0.02)$ \\  
  $B^{*+}$           &  $-(0.80 \pm 0.10)$ \\  
  $B^{*0}$           &  $-(0.20 \pm 0.03)$ \\  
  $B_{s}^{*0}$       &  $-(0.17 \pm 0.03)$ \\
  $D_{1}^{0}$        &  $(0.18 \pm 0.02)$  \\ 
  $D_{1}^{+}$        &  $-(0.60 \pm 0.02)$  \\
  $D_{s1}$           &  $-(0.59 \pm 0.02)$  \\
  $B_{1}^{0}$        &  $-(0.12 \pm 0.02)$  \\
  $B_{1}^{+}$        &  $-(0.78 \pm 0.02)$ \\
  $B_{s1}$           &  $-(0.10 \pm 0.02)$ \\      
 \bottomrule
    \end{tabular}
    \caption{The quadrupole moments of heavy vector and axial vector mesons are depicted in natural units.}
    \label{tab:5}
\end{table}

We see that the values of the magnetic moments of $D^{* +}$ and $D_s^{*+}$ predicted by the light cone sum rules framework are in good agreement with the other approaches. Once the uncertainties are taken into account, our result on $B_s^{*0}$ magnetic moment is compatible with the prediction of the chiral perturbation theory.

$SU(3)$ symmetry dictates that, the magnetic moments of $D^{*+}$, $D_s^{+}$, $B^{*0}$, $B_s^{0}$,$D_{s1}$, $D_1^+$, $B_1^{0}$, and $B_{s1}^0$ should be very close to each other. Our predictions for the magnetic moments of these mesons are in good agreement  within $SU(3)$ symmetry expectations. The violation of $SU(3)$ symmetry is about maximum $20\%$. The violation of $SU(3)$ symmetry is due to the mass of strange quark, the different values of quark condensates for $u$, $d$, and $s$ quarks as well as the values of the leptonic decay constants. However, predicitions of chiral perturbation theory~\cite{PhysRevD.100.016019} leads huge ($4$-times) violation of $SU(3)$ symmetry which seems highly unnatural.

The difference between our predictions with the other approaches on the magnetic moments can be explained as follows. The main contribution to the magnetic moments in light-cone sum rules results from the perturbative part of the spectral density. The perturbative part schematically can be written as
\begin{equation}
  \label{eq:15}
  (e_Q - e_q)A + e_Q B
\end{equation}
where numerically $A$ is larger than $B$. In the charged meson case, $e_Q - e_q$ is equal to one and for this reason magnetic moment is quite large. However, for the neutral meson case, $e_Q - e_q = 0$, and hence, the magnetic moment is rather small. Our last remark to this section is as follows. To increase the precision of our calculations, the next-to-leading order (NLO) QCD corrections to the correlation functions should be taken into account. In addition, the same order of NLO corrections should be included for the calculation of the leptonic decay constants. However, since the expressions for considered multipole moments depend on the ratio of these two factors, it is expected that our findings may not be changed considerably. 

Finally,  we would like to emphasize that the magnetic moments of axial-vector heavy mesons are calculated first time to our knowledge. It would be interesting to have results within  other approaches for the magnetic moments of these mesons.

\section{Conclusion}
The magnetic and quadrupole moments of vector and axial-vector mesons containing heavy quark are estimated within the light cone QCD sum rules framework by using photon distribution amplitudes. The magnetic moments of axial-vector mesons are estimated  for the first time. Moreover, we compared our predictions on magnetic moments with the results obtained from other approaches. Our findings agree with the results of other methods for $D^{*+}$ and $D_s^*$ mesons. Besides, our predictions for the magnetic moments of $D^{*+}$, $D_s^{+}$, $B^{*0}$, $B_s^{0}$,$D_{s1}$, $D_1^+$, $B_1^{0}$, and $B_{s1}^0$ mesons are consistent with $SU(3)$ symmetry expectations. However, chiral perturbation theory predicts huge $SU(3)$ symmetry breaking (about four times), which is bizarre. The calculation of these magnetic moments for the axial-vector mesons within other approaches can be very useful for understanding the inner structure of heavy mesons. 

\clearpage
\newpage

\appendix
\section{Photon Distribution Amplitudes}
In this appendix, we present the matrix element of non-local operators in terms of photon distribution amplitudes (DAs) and their explicit expressions.


\begin{equation*}
  \label{eq:19}
  \begin{split}
 \lla \gamma(q) \vert  \bar q(x) \sigma_{\mu \nu} q(0) \vert  0 \rra  &=
-i e_q \qq (\varepsilon_\mu q_\nu - \varepsilon_\nu q_\mu) 
\int_0^1 du e^{i \bar u qx} 
\Bigg(\chi \varphi_\gamma(u) + {x^2 \over 16} \mathbb{A}  (u) \Bigg) \\  
& - {i\over 2(qx)}  e_q \qq \Bigg[x_\nu \Bigg(\varepsilon_\mu - 
q_\mu {\varepsilon x \over qx}\Bigg) - 
x_\mu \Bigg(\varepsilon_\nu - q_\nu {\varepsilon x\over q x}\Bigg) \Bigg] \\
& ~~~~~~~~~~~~~~\times
\int_0^1 du e^{i \bar u q x} h_\gamma(u)  \\
\lla \gamma(q) \vert  \bar q(x) \gamma_\mu q(0) \vert 0 \rra  &=
e_q f_{3 \gamma} \Bigg(\varepsilon_\mu - q_\mu {\varepsilon x\over q x} 
\Bigg) \int_0^1 du e^{i \bar u q x} \psi^v(u) \\
   \lla \gamma(q) \vert \bar q(x) \gamma_\mu \gamma_5 q(0) \vert 0 \rra  &=
- {1\over 4} e_q f_{3 \gamma} \epsilon_{\mu \nu \alpha \beta } 
\varepsilon^\nu q^\alpha x^\beta \int_0^1 du e^{i \bar u q x} 
\psi^a(u) \\
 \lla \gamma(q) | \bar q(x) g_s G_{\mu \nu} (v x) q(0) \vert 0 \rra &=
-i e_q \qq \Bigg(\varepsilon_\mu q_\nu - \varepsilon_\nu q_\mu \Bigg) 
\int {\cal D}\alpha_i e^{i (\alpha_{\bar q} + v \alpha_g) q x} 
{\cal S}(\alpha_i) \\
\lla \gamma(q) | \bar q(x) g_s \tilde G_{\mu \nu} i \gamma_5 (v x) 
q(0) \vert 0 \rra &= -i e_q \qq \Bigg(\varepsilon_\mu q_\nu - 
\varepsilon_\nu q_\mu \Bigg)  \int {\cal D}\alpha_i e^{i (\alpha_{\bar q} + 
  v \alpha_g) q x} \tilde {\cal S}(\alpha_i)  \\
\lla \gamma(q) \vert \bar q(x) g_s \tilde G_{\mu \nu}(v x) 
\gamma_\alpha \gamma_5 q(0)  \vert 0 \rra &=
e_q f_{3 \gamma} q_\alpha (\varepsilon_\mu q_\nu - \varepsilon_\nu q_\mu) 
\int {\cal D}\alpha_i e^{i (\alpha_{\bar q} + v \alpha_g) q x} {\cal A}
(\alpha_i) \\
 \lla \gamma(q) \vert \bar q(x) g_s G_{\mu \nu}(v x) i \gamma_\alpha q(0) 
\vert 0 \rra &= e_q f_{3 \gamma} q_\alpha (\varepsilon_\mu q_\nu - 
\varepsilon_\nu q_\mu) \int {\cal D}\alpha_i e^{i (\alpha_{\bar q} + 
  v \alpha_g) q x} {\cal V}(\alpha_i)  \\
  \end{split}
\end{equation*}

\begin{equation*}
  \begin{split}
 \lla \gamma(q) \vert \bar q(x) \sigma_{\alpha \beta} g_s 
G_{\mu \nu}(v x) q(0) \vert 0 \rra  &= e_q \qq \Bigg\{
\Bigg[\Bigg(\varepsilon_\mu - q_\mu {\varepsilon x\over q x}\Bigg)
\Bigg(g_{\alpha \nu} - {1\over qx} (q_\alpha x_\nu + q_\nu x_\alpha)
\Bigg) q_\beta  \\
 &- \Bigg(\varepsilon_\mu - q_\mu {\varepsilon x\over q x}\Bigg)
\Bigg(g_{\beta \nu} - {1\over qx} (q_\beta x_\nu + q_\nu x_\beta)
\Bigg) q_\alpha  \\
& - \Bigg(\varepsilon_\nu - q_\nu {\varepsilon x\over q x}\Bigg)
\Bigg(g_{\alpha \mu} - {1\over qx} (q_\alpha x_\mu + q_\mu x_\alpha)
\Bigg) q_\beta  \\
& + \Bigg(\varepsilon_\nu - q_\nu {\varepsilon 
x\over qx}\Bigg)\Bigg( g_{\beta \mu} - {1\over qx} (q_\beta x_\mu + 
q_\mu x_\beta)\Bigg) q_\alpha \Bigg] \\
&~~~~~~~~~~~~ \times \int {\cal D}\alpha_i 
e^{i (\alpha_{\bar q} + v \alpha_g) qx} {\cal T}_1(\alpha_i) \\
& + \Bigg[\Bigg(\varepsilon_\alpha - q_\alpha {\varepsilon x\over 
qx}\Bigg) \Bigg(g_{\mu \beta} - {1\over qx}(q_\mu x_\beta + q_\beta 
x_\mu)\Bigg)  q_\nu  \\
& - \Bigg(\varepsilon_\alpha - q_\alpha {\varepsilon x\over qx}
\Bigg) \Bigg(g_{\nu \beta} - {1\over qx}(q_\nu x_\beta + q_\beta 
x_\nu)\Bigg)  q_\mu  \\
& - \Bigg(\varepsilon_\beta - q_\beta {\varepsilon x\over qx}\Bigg)
\Bigg(g_{\mu \alpha} - {1\over qx}(q_\mu x_\alpha + q_\alpha 
x_\mu)\Bigg) q_\nu  \\
& + \Bigg(\varepsilon_\beta - q_\beta {\varepsilon x\over qx}\Bigg)
\Bigg(g_{\nu \alpha} - {1\over qx}(q_\nu x_\alpha + q_\alpha x_\nu) 
\Bigg) q_\mu \Bigg] \\
& ~~~~~~~~~~~ \times \int {\cal D} \alpha_i e^{i (\alpha_{\bar q} + 
  v \alpha_g) qx} {\cal T}_2(\alpha_i)  \\
& + {1\over qx} (q_\mu x_\nu - q_\nu x_\mu) (\varepsilon_\alpha 
q_\beta - \varepsilon_\beta q_\alpha) \int {\cal D} \alpha_i 
e^{i (\alpha_{\bar q} + v \alpha_g) qx} {\cal T}_3(\alpha_i) \\
& + {1\over qx} (q_\alpha x_\beta - q_\beta x_\alpha)
(\varepsilon_\mu q_\nu - \varepsilon_\nu q_\mu)
\int {\cal D} \alpha_i e^{i (\alpha_{\bar q} + v \alpha_g) qx} 
{\cal T}_4(\alpha_i) \Bigg\}~,        
  \end{split}
\end{equation*}
where $\chi$ is the magnetic susceptibility of the quarks, $\varphi_\gamma(u)$ is the leading
twist 2, $\psi^v(u)$, $\psi^a(u)$, ${\cal A}$ and ${\cal V}$ are the twist 3 and $h_\gamma(u)$,
$\mathbb{A}$, ${\cal T}_i$ ($i=1,~2,~3,~4$) are the twist 4 photon distribution amplitudes. The
integral measure ${\cal D} \alpha_i $ is defined as
\bea
\label{e9512}
{\cal D} \alpha_i = \int_0^1 d\alpha_g \int_0^1 d\alpha_q \int_0^1
d\alpha_{\bar{q}}~ \delta(1-\alpha_g - \alpha_q  - \alpha_{\bar{q}})~.
\eea
The expressions of the photon DAs which we need in our calculations are~\cite{Ball:2002ps}:
\bea
\label{eesr16}
\varphi_\gamma(u) \es 6 u \bar u \Big[ 1 + \varphi_2(\mu)
C_2^{\frac{3}{2}}(u - \bar u) \Big]~,
\nnb \\
\psi^v(u) \es 3 [3 (2 u - 1)^2 -1 ]+\frac{3}{64} (15
w^V_\gamma - 5 w^A_\gamma)
                        [3 - 30 (2 u - 1)^2 + 35 (2 u -1)^4]~,
\nnb \\
\psi^a(u) \es [1- (2 u -1)^2] [ 5 (2 u -1)^2 -1 ]
\frac{5}{2}
    \Bigg(1 + \frac{9}{16} w^V_\gamma - \frac{3}{16} w^A_\gamma
    \Bigg)~,
\nnb \\
{\cal A}(\alpha_i) \es 360 \alpha_q \alpha_{\bar q} \alpha_g^2
        \Bigg[ 1 + w^A_\gamma \frac{1}{2} (7 \alpha_g - 3)\Bigg]~,
\nnb \\
{\cal V}(\alpha_i) \es 540 w^V_\gamma (\alpha_q - \alpha_{\bar q})
\alpha_q \alpha_{\bar q}
                \alpha_g^2~,
\nnb \\
h_\gamma(u) \es - 10 (1 + 2 \kappa^+ ) C_2^{\frac{1}{2}}(u
- \bar u)~,
\nnb \\
\mathbb{A}(u) \es 40 u^2 \bar u^2 (3 \kappa - \kappa^+ +1 ) +
        8 (\zeta_2^+ - 3 \zeta_2) [u \bar u (2 + 13 u \bar u) + 
                2 u^3 (10 -15 u + 6 u^2) \ln(u) \nnb \\ 
\ar 2 \bar u^3 (10 - 15 \bar u + 6 \bar u^2)
        \ln(\bar u) ]~,
\nnb \\
{\cal T}_1(\alpha_i) \es -120 (3 \zeta_2 + \zeta_2^+)(\alpha_{\bar
q} - \alpha_q)
        \alpha_{\bar q} \alpha_q \alpha_g~,
\nnb \\
{\cal T}_2(\alpha_i) \es 30 \alpha_g^2 (\alpha_{\bar q} - \alpha_q)
    [(\kappa - \kappa^+) + (\zeta_1 - \zeta_1^+)(1 - 2\alpha_g) +
    \zeta_2 (3 - 4 \alpha_g)]~,
\nnb \\
{\cal T}_3(\alpha_i) \es - 120 (3 \zeta_2 - \zeta_2^+)(\alpha_{\bar
q} -\alpha_q)
        \alpha_{\bar q} \alpha_q \alpha_g~,
\nnb \\
{\cal T}_4(\alpha_i) \es 30 \alpha_g^2 (\alpha_{\bar q} - \alpha_q)
    [(\kappa + \kappa^+) + (\zeta_1 + \zeta_1^+)(1 - 2\alpha_g) +
    \zeta_2 (3 - 4 \alpha_g)]~,\nnb \\
{\cal S}(\alpha_i) \es 30\alpha_g^2\{(\kappa +
\kappa^+)(1-\alpha_g)+(\zeta_1 + \zeta_1^+)(1 - \alpha_g)(1 -
2\alpha_g)\nnb \\ 
\ar\zeta_2
[3 (\alpha_{\bar q} - \alpha_q)^2-\alpha_g(1 - \alpha_g)]\}~,\nnb \\
\tilde {\cal S}(\alpha_i) \es-30\alpha_g^2\{(\kappa -
\kappa^+)(1-\alpha_g)+(\zeta_1 - \zeta_1^+)(1 - \alpha_g)(1 -
2\alpha_g)\nnb \\ 
\ar\zeta_2 [3 (\alpha_{\bar q} -
\alpha_q)^2-\alpha_g(1 - \alpha_g)]\},
\eea
where $C_n^m$ is the Gegenbauer polynomial. The constants entering  the above DAs are adapted from~\cite{Ball:2002ps} and their values are given in Table~\ref{tab:2}.

\begin{table}
   \renewcommand{\arraystretch}{1.05}
\setlength{\tabcolsep}{10pt}
  \centering
  \begin{tabular}{cccccccccc}
    \toprule
    $\varphi_2$ & $ \kappa $ & $\kappa^+$ & $\xi_1$ & $\xi_1^+$ & $\xi_2$ & $\xi_2^+$ & $f_{3\gamma}~(GeV^2)$          & $\omega_\gamma^V$ & $\omega_\gamma^A$ \\
    \midrule
   $0.0$      & $0.2$      & $0.0$      & $0.4$   & $0.0$     & $0.3$   & $0.0$     & $(-4.0 \pm 2.0 ) \times10^{-3}$ & $(3.8 \pm 1.8$)   & $(-2.1 \pm 1.0)$  \\
\bottomrule
  \end{tabular}
  \caption{The values of the constant parameters entering into
the distribution amplitudes at the renormalization scale $\mu=1~GeV$.}
  \label{tab:2}
\end{table}

\section{Explicit expressions of the invariant functions}
In this Appendix, we present the explicit expressions of the invariant functions $\Pi_1^{(\pm)}$, $\Pi_2^{(\pm)}$, and $\Pi_3^{(\pm)}$.  
\subsubsection{Coefficient of the $(\varepsilon\!\cdot\!p) p_{\mu} p_{\nu}$ structure}
\begin{equation}
  \label{eq:16}
  \begin{split}
\Pi_1^{(\pm)} &=~ - {1\over \pi^2} \Big[3 (e_Q - e_q) m_Q^4 M^2 ({\cal I}_3 - m_Q^2 {\cal I}_4)\Big]  \\
&-  {e^{-m_b^2/M^2} \over 24 \pi^2 M^2} \Big[48 \pi^2 e_Q m_q \langle \bar{q}q \rangle 
+ e^{m_b^2/M^2} e_q \langle g_s^2 G^2 \rangle m_Q^2 
{\cal I}_2)\Big]  \\
&+  {e^{-m_b^2/M^2}\over 3 M^4} e_Q m_0^2 m_q \langle \bar{q}q \rangle 
+ {e^{-m_b^2/M^2}\over 3 M^6} e_Q m_0^2 m_Q^2 m_q 
\langle \bar{q}q \rangle~.
  \end{split}
\end{equation}
\subsubsection{Coefficient of the $(\varepsilon\!\cdot\!p) p_{\nu} q_{\mu}$ structure}
\begin{equation}
  \label{eq:17}
  \begin{split}
\Pi_2^{(\pm)}~ \!\!\! &=~ -{3\over 4 \pi^2} m_Q^2 M^2 \Big[e_Q {\cal I}_2 + (e_Q - 3 e_q) m_Q^2 {\cal I}_3 - 
   2 (e_Q - e_q) m_Q^4 {\cal I}_4 \Big] \\
&- {e^{-m_b^2/M^2} \over 96 \pi^2 M^2}
\Big[e_q \langle g_s^2 G^2 \rangle +144 e_Q m_q \pi^2 \langle \bar{q}q \rangle + 
2 e^{m_b^2/M^2} e_q \langle g_s^2 G^2 \rangle m_Q^2 {\cal I}_2 \Big] \\
&\pm  {e^{-m_b^2/M^2} \over M^2} \Big[2 e_q m_Q \langle \bar{q}q \rangle 
\widetilde{j}_2(h_\gamma)\Big] + {e^{-m_b^2/M^2} \over 6 M^4} e_Q m_0^2 m_q \langle \bar{q}q \rangle  \\
&+ {e^{-m_b^2/M^2} \over 144 M^6}
m_Q \Big\{36 e_Q m_0^2 m_Q m_q \langle \bar{q}q \rangle
+ e_q \langle g_s^2 G^2 \rangle \Big[
\pm 12 \langle \bar{q}q \rangle \widetilde{j}_2(h_\gamma)  \\
&+ f_{3\gamma} m_Q \Big(4 \widetilde{j}_1(\psi^v) +
\psi^a(u_0)\Big)\Big]\Big\}  \\
&\mp  {e^{-m_b^2/M^2} \over 36 M^8} e_q \langle g_s^2 G^2 \rangle m_Q^3 \langle \bar{q}q \rangle 
\widetilde{j}_2(h_\gamma)  \\
&-  {e^{-m_b^2/M^2} \over 2} e_q f_{3\gamma} \Big[4 \widetilde{j}_1(\psi^v) + 
\psi^a(u_0)\Big]~. 
  \end{split}
\end{equation}
\subsubsection{Coefficient of the $(\varepsilon\!\cdot\!p) q_{\mu} q_{\nu}$ structure}
\begin{equation}
  \label{eq:18}
  \begin{split}
    \Pi_3^{(\pm)}~ \!\!\! &=~ - {1\over 4 \pi^2}
\Big[3 (e_Q - e_q) m_Q^4 M^2 ({\cal I}_3 - m_Q^2 {\cal I}_4)\Big] \\
&-  {e^{-m_b^2/M^2} \over 96 M^2 \pi^2} \Big\{48 e_Q m_q \pi^2 \langle \bar{q}q \rangle + e^{m_b^2/M^2} e_q \langle g_s^2 G^2 \rangle m_Q^2 {\cal I}_2 \\
&\pm 384 e_q m_Q \pi^2 \langle \bar{q}q \rangle \Big[i_1^\prime({\cal T}_1,1)
+ i_1^\prime({\cal T}_2,1) - 
i_1^\prime({\cal T}_3,1) - i_1^\prime({\cal T}_4,1)\Big]\Big\} \\ 
&+ {e^{-m_b^2/M^2} \over 12 M^4}
e_Q m_0^2 m_q \langle \bar{q}q \rangle
+ {e^{-m_b^2/M^2} \over 36 M^6}
\Big\{m_Q^2 \Big[3 e_Q m_0^2 m_q \langle \bar{q}q \rangle + e_q f_{3\gamma} \langle g_s^2 G^2 \rangle
\widetilde{j}_1(\psi^v)\Big]\Big\}  \\
&+ e^{-m_b^2/M^2}\Big\{2 e_q f_{3\gamma} \Big[2 i_2({\cal V},v) +
i_2^\prime({\cal A},1) - i_2^\prime({\cal V},1) -
\widetilde{j}_1(\psi^v)\Big]\Big\}~.
  \end{split}
\end{equation}
where
\begin{eqnarray}
\label{nolabel}
{\cal I}_n~ \!\!\! &=~& \!\!\! \int_{m_b^2}^{s_0} ds\, 
{e^{-s/M^2} \over s^n}~,\nonumber \\
i_1(\phi,f(v)) \!\!\! &~=~& \!\!\! \int {\cal D}\alpha_i \int_0^1 dv
\phi(\alpha_{\bar{q}},\alpha_q,\alpha_g) f(v) \theta(k-u_0)~, \nonumber \\
i_2(\phi,f(v)) \!\!\! &~=~& \!\!\! \int {\cal D}\alpha_i \int_0^1 dv
\phi(\alpha_{\bar{q}},\alpha_q,\alpha_g) f(v) \delta(k-u_0)~, \nonumber \\
i_1^\prime(\phi,f(v)) \!\!\! &~=~& \!\!\! \int {\cal D}\alpha_i \int_0^1 dv
\phi(\alpha_{\bar{q}},\alpha_q,\alpha_g) f(v) \theta(k^\prime-u_0)~, \nonumber \\
i_2^\prime(\phi,f(v)) \!\!\! &~=~& \!\!\! \int {\cal D}\alpha_i \int_0^1 dv
\phi(\alpha_{\bar{q}},\alpha_q,\alpha_g) f(v) \delta(k^\prime-u_0)~, \nonumber \\
\widetilde{j}_1(f(u)) \!\!\! &~=~& \!\!\! \int_{u_0}^1 du f(u)~, \nonumber \\
\widetilde{j}_2(f(u)) \!\!\! &~=~& \!\!\! \int_{u_0}^1 du (u-u_0) f(u)~, \nonumber
\end{eqnarray}
and,
\begin{eqnarray}
k = \alpha_q + \alpha_g v,~k^\prime = \alpha_q + \alpha_g (1-v),
~M^2={M_1^2 M_2^2 \over M_1^2 +M_2^2}~,
~u_0={M_1^2 \over M_1^2 +M_2^2}. \nonumber
\end{eqnarray}

In calculations, we take $M_1^2 = M_2^2$ since the initial and final state mesons are same, hence $u_0 = \frac{1}{2}$.


\bibliographystyle{apsrev4-1}
\bibliography{multipole}

\end{document}